\documentclass{article}
\usepackage{slashed,feynmf,epsfig,amsmath,amssymb,enumitem}
\usepackage[papersize={8.5in,11in}]{geometry}
\begin{document}
\title{Superluminal Gravitational Waves}
\author{J. W. Moffat\\~\\
Perimeter Institute for Theoretical Physics, Waterloo, Ontario N2L 2Y5, Canada\\
and\\
Department of Physics and Astronomy, University of Waterloo, Waterloo,\\
Ontario N2L 3G1, Canada}
\maketitle




\begin{abstract}
The quantum gravity effects of vacuum polarization of gravitons propagating in a curved spacetime cause the quantum vacuum to act as a dispersive medium with a refractive index. Due to this dispersive medium gravitons acquire superluminal velocities. The dispersive medium is produced by higher derivative curvature contributions to the effective gravitational action. It is shown that in a Friedmann-Lema\^{i}tre-Robertson-Walker spacetime in the early universe near the Planck time $t_{\rm PL}\gtrsim 10^{-43}\,{\rm sec}$, the speed of gravitational waves $c_g\gg c_{g0}=c_0$, where $c_{g0}$ and $c_0$ are the speeds of gravitational waves and light today. The large speed of gravitational waves stretches their wavelengths to super-horizon sizes, allowing them to be observed in B-polarization experiments.
\end{abstract}

\maketitle

\section{Introduction}

A variable speed of light cosmology has been developed~\cite{Moffat,Moffat1,Moffat2} which solves the initial value problems in early universe cosmology: the horizon problem and the flatness problem. The premise is that fractions of seconds after the big bang the speed of light $c\gg c_0$, where $c_0$ is the measured speed of light today. When in the same epoch the speed of gravitational waves $c_g\gg c_{g0}$ where $c_{g0}$ is the speed of gravitational waves today and $c_{g0}=c_0$, then the wavelengths of sub-horizon tiny quantum perturbative fluctuations $\delta\phi$ and primordial gravitational waves are stretched super-horizon to sizes that allow their detection in the radiation emitted from the CMB. A calculation of the curvature power spectra for $\delta\phi$ and the primordial gravitational waves yields the spectral indices (tilts), $n_s=0.96$ in good agreement with the Planck Mission result $n_s= 0.9603\pm 0.0073$~\cite{PlanckMission1,PlanckMission2,PlanckMission3} and $n_t=-0.04$.  Adopting the BICEP2 B-polarization result $r=0.2$ (for zero or small foreground dust contamination)~\cite{BICEP2}, then $r/n_t=-5$ which is close to the single-field inflation self-consistency condition $r/n_t=-8$. Homogeneous Lorentz invariance is spontaneously violated in the phase when $c\gg c_0$ and $c_g\gg c_{g0}$, so that $SO(3,1)\rightarrow O(3)$. A preferred proper comoving time $t$ is chosen in the Lorentz symmetry violating phase corresponding to the comoving time in a Friedmann-Lema\^{i}tre-Robertson-Walker (FLRW) spacetime with the symmetry $O(3)\times R$. 

A fundamental explanation for the origin of the large speed of light in the early universe is provided by the Drummond-Hathrell calculation of the speed of light propagation in the dispersive medium with a refractive index, caused by the quantum electrodynamic (QED) vacuum polarization in a curved spacetime~\cite{Drummond,Shore1,Shore2}. A non-minimal coupling of the electromagnetic field to the curvature tensor produces quantum corrections and the photon acquires a size and an effective mass from the tidal gravitational forces on the photon. In the following, we will derive a similar result that explains the superluminal speed of gravitational waves in the very early universe, caused by a non-minimal coupling of the graviton to higher derivative curvature and the resulting tidal forces on the graviton. In the early universe near the Planck energy, the gravitational fine structure constant $\alpha_g=Gm^2/\hbar c_0\sim 1$, allowing for sufficiently large quantum gravity (QG) vacuum polarization comparable to the photon vacuum polarization obtained from the QED fine structure coupling strength $\alpha=e^2/\hbar c_0$. This vacuum polarization induces a dispersive medium with a refractive index that causes the speed of gravitational waves to be large in the early universe near the Planck energy. The wavelengths of primordial gravitational waves, $\lambda_g=c_g/\nu_g$, where $\nu_g$ is the frequency of the waves, are stretched super-horizon allowing the primordial waves to be detected by B-polarization experiments.

\section{Graviton propagation}

We will consider graviton propagation in a curved spacetime determined by the effective action:
\begin{equation}
\label{EffecGrav}
{\cal L}_{\rm effG}=\frac{1}{\kappa}\int d^4x\sqrt{-g}[R+\lambda_c^2(\gamma_1R^2+\gamma_2R_{\mu\nu}R^{\mu\nu}+\gamma_3R_{\mu\nu\rho\sigma}R^{\mu\nu\rho\sigma})],
\end{equation}
where $\kappa=16\pi G/c_0^4$ is constant throughout the evolution of the universe, $\gamma_1,\gamma_2$ and $\gamma_3$ are dimensionless coefficients and $\lambda_c=\hbar/mc_0$ is the Compton wavelength for a mass $m$. We will ignore possible contributions from higher derivative curvature terms. 

For the case of the action with $\gamma_1=\gamma_2=\gamma_3=0$, we have the vacuum equation of motion:
\begin{equation}
R_{\mu\nu}=0.
\end{equation}
By expanding the metric $g_{\mu\nu}=\eta_{\mu\nu}+h_{\mu\nu}$, where $\eta_{\mu\nu}$ is the Minkowski spacetime metric, the vacuum field equations to lowest order in the weak field 
approximation are given by
\begin{equation}
\label{graveq1}
\eta^{\alpha\beta}\partial_\alpha\partial_\beta h_{\mu\nu}=0,
\end{equation}
provided we impose the harmonic coordinate condition~\cite{Weinberg}:
\begin{equation}
\label{graveq2}
\partial_\mu {h^\mu}_\nu=\frac{1}{2}\partial_\nu {h^\mu}_\mu.
\end{equation}

We will determine the characteristics of graviton propagation using a geometric optics approximation. In the leading order, the perturbative metric $h_{\mu\nu}$ is written as a slowly varying amplitude and a rapidly varying phase:
\begin{equation}
h_{\mu\nu}(x)=a_{\mu\nu}\exp\theta(x),
\end{equation}
where $a_{\mu\nu}$ is a polarization tensor and the graviton wave vector is defined by $k_\mu=\partial_\mu\theta$, where in the QG interpretation in terms of gravitons, $k_\mu$ is identified as the graviton momentum. Coordinate derivatives act only on the phase $\theta$. 

In terms of the graviton momentum $k^\mu$ the equations (\ref{graveq1}) and (\ref{graveq2}) become
\begin{equation}
\label{Gravwaveeq}
k^2 a_{\mu\nu}=0,
\end{equation}
and
\begin{equation}
k_\mu {a^\mu}_\nu=\frac{1}{2}k_\nu{a^\mu}_\mu. 
\end{equation}
From (\ref{Gravwaveeq}) we have $k^2=k^\mu k_\mu=0$ and $k^\mu$ is a null vector. It follows from the definition of $k_\mu$ as a gradient $k_\mu=\partial_\mu\theta$ that $\nabla_\mu k^\nu=\nabla^\nu k_\mu$ whereby 
\begin{equation}
\label{Momentumeq}
k^\mu\nabla_\mu k^\nu=k^\mu\nabla^\nu k_\mu=\frac{1}{2}\nabla^\nu k^2=0.
\end{equation}
Graviton trajectories are defined as the curves of the wave vector $k^\mu$ for which $k^\mu=dx^\mu/ds$ where $s$ is the proper time along the trajectory. Substituting into (\ref{Momentumeq}), we get the null geodesic equation for the graviton:
\begin{equation}
k^\mu\nabla_\mu k^\nu=\frac{d^2x^\nu}{ds^2}+{\Gamma^\nu}_{\mu\lambda}\frac{dx^\mu}{ds}\frac{dx^\lambda}{ds}=0.
\end{equation}

These properties of gravitons are no longer true when we consider the equations of motion obtained from the non-minimal effective gravitational action (\ref{EffecGrav}) including QG corrections. The equations of motion now become
\begin{equation}
R_{\mu\nu}+\lambda_c^2[\gamma_1P_{\mu\nu}(R^2)+\gamma_2L_{\mu\nu}(R_{\mu\nu}^2)+\gamma_3K_{\mu\nu}(R_{\mu\nu\rho\sigma}^2)]=0,
\end{equation}
where $P_{\mu\nu}, L_{\mu\nu}$ and $K_{\mu\nu}$ are the contributions from the non-minimal coupling to $R^2$, $R_{\mu\nu}^2$ and $R_{\mu\nu\rho\sigma}^2$, respectively. In the momentum phase space geometric optics approximation and for weak gravitational radiation this becomes
\begin{equation}
\label{ModConeeq}
k^2a_{\mu\nu}+\lambda_c^2[\gamma_1P_{\mu\nu}(k^2,a) +\gamma_2L_{\mu\nu}(k^2,a)+\gamma_3K_{\mu\nu}(k^2,a)]=0.
\end{equation}
We choose the polarization tensor $a_{\mu\nu}$ to be spacelike normalized, $a^{\mu\nu}a_{\mu\nu}=-1$. Contracting (\ref{ModConeeq}) with $a^{\mu\nu}$ we find
\begin{equation}
\label{LightConeeq}
k^2 - \lambda_c^2[\gamma_1P(k^2,a)+\gamma_2L(k^2,a)+\gamma_3K(k^2,a)]=0,
\end{equation}
where $P=a^{\mu\nu}P_{\mu\nu}$, $L=a^{\mu\nu}L_{\mu\nu}$ and $K=a^{\mu\nu}K_{\mu\nu}$. This equation contains position dependent curvature terms, which means that it does not reduce to the special relativistic equations at the origin of inertial frames of reference, for different inertial frames differ at different points of spacetime. The equation violates the 
Strong Equivalence Principle, which states that all inertial frames are equivalent. This coincides with our spontaneous symmetry breaking of Lorentz invariance invoked in our VSL model~\cite{Moffat}.  The violation of causality and the generation of closed time-like curves depends on the two conditions being met: (1) spacelike photon and graviton curves and, (2) Lorentz invariance. When the Strong Equivalence Principle and Lorentz invariance are violated, then causality may still be retained. 

Because the modification of the light cone by (\ref{LightConeeq}) is homogeneous in $k^\mu$, we can express the equation in a new metric:
\begin{equation}
{\tilde g}^{\mu\nu}k_\mu k_\nu=0,
\end{equation}
where ${\tilde g}^{\mu\nu}$ takes into account the curvature of spacetime and the polarization of the graviton. We can define a new momentum 4-vector:
\begin{equation}
q^\mu={\tilde g}^{\mu\nu}k_\nu,
\end{equation}
and we have the equation:
\begin{equation}
{\hat g}_{\mu\nu}q^\mu q^\nu={\tilde g}^{\mu\nu}k_\mu k_\nu=0,
\end{equation}
where ${\hat g}_{\mu\nu}=({\tilde g}_{\mu\nu})^{-1}$ defines a new metric and we raise and lower indices for $q^\mu$ with $g_{\mu\nu}$. We can describe the propagation of photons and gravitons in a quantized vacuum as a bimetric theory in which the physical light cones are determined by the effective metric ${\tilde g}_{\mu\nu}$, while the geometric null cones are fixed by the spacetime metric $g_{\mu\nu}$.

\section{Propagation of gravitons in cosmology}

We will work in our application to cosmology in a frame in which the metric is of the FLRW form:
\begin{equation}
\label{FLRW}
ds^2=c^2dt^2-a^2\biggl[\frac{dr^2}{1-Kr^2}+r^2(d\theta^2+\sin^2\theta d\phi^2)\biggr],
\end{equation}
with $K=0,+1,-1$ for flat, closed and open models. The metric has the group symmetry $O(3)\times R$ with a preferred proper comoving time $t$.
The energy-momentum tensor $T_{\mu\nu}$ will be described by a perfect fluid:
\begin{equation}
T^{\mu\nu}=\biggl(\rho+P\biggr)u^\mu u^\nu-Pg^{\mu\nu},
\end{equation}
where $u^\mu=dx^\mu/d\tau$ is the fluid element four-velocity, $d\tau=ds/c_0$, $P=p/c_0^2$ and $\rho$ and $p$ are the matter density and pressure, respectively. The four-velocity $u^\mu$ has the comoving frame values $u^0=1$ and $u^i=0\,(i=1,2,3)$.

The FLRW spacetime is Weyl curvature flat, homogeneous and isotropic, so the light cone equation (\ref{LightConeeq}) takes the form:
\begin{equation}
k^2=\zeta^2_g8\pi G(\rho+P)(k^\mu u_\mu)^2,
\end{equation}
where 
\begin{equation}
\zeta^2_g=\lambda_c^2(d\gamma_1+e\gamma_2+f\gamma_3),
\end{equation}
and the coefficients $d,e,f$ are to be determined by the QG calculation of vacuum polarization loops. The vacuum polarization at one-loop order corresponds to the Feynman graph connecting a graviton  to another graviton through a fermion-anti-fermion loop with the coupling strength $\alpha_g=Gm^2/\hbar c_0$. 

In the early universe in the radiation dominated era $a\propto t^{1/2}$, $\rho\propto 1/t^2$ and with $k^2=(k^0)^2-\vert{\bf k}\vert^2$ and $(k^\mu u_\mu)^2=(k^0)^2$ we get 
\begin{equation}
\frac{c^2_g}{c_{g0}^2}=\frac{(k^0)^2}{\vert{\bf k}\vert^2}=\frac{1}{1-t_{g*}^2/t^2},
\end{equation}
where $t_{g*}=\zeta_g/c_0$. This can be rewritten as
\begin{equation}
c_g=\biggl(1-\frac{t_{g*}^2}{t^2}\biggr)^{-1/2}c_{g0}.
\end{equation}
This is equivalent to the Drummond-Hathrell result for photon propagation in the early universe~\cite{Drummond,Shore1,Shore2,Moffat}:
\begin{equation}
c=\biggl(1-\frac{11}{2}\frac{t_*^2}{t^2}\biggr)^{-1/2}c_0,
\end{equation}
where $t_*=\xi/c_0$, $\xi=\biggl(\alpha/90\pi\biggr)^{1/2}\lambda_e$, and where $\alpha=e^2/\hbar c_0$ and $\lambda_e=\hbar/m_ec_0$ are the QED fine structure constant and the Compton wavelength of the electron, respectively. We have $c\gg c_0$ for $t\gtrsim \sqrt{11/2}t_*=2\times 10^{-23}\,{\rm sec}$ after the big bang, corresponding to an energy $\,\sim 2\times 10^5$ TeV. We expect that $c_g\gg c_{g0}$ will occur for $t\gtrsim t_{g*}$ much earlier in the universe than the large increase in the speed of light $c$, namely, near the the Planck energy $mc_0^2=(\hbar c_0^5/G)^{1/2}$ and the Planck time $t_{\rm PL}\sim 10^{-43}\,{\rm sec}$. Near the Planck energy the gravitational fine structure constant $\alpha_g\sim {\cal O}(1)$ and it becomes comparable to the fine structure constant $\alpha\sim 1/137$, so that the QG vacuum polarization calculation will produce a dispersive medium with a refractive index $n_{\rm grefrac}$ that allows for $c_g\gg c_{g0}=c_0$. Once the primordial gravitational waves have been stretched 
super-horizon, $\lambda_g=c_g/\nu_g$, and ``frozen'' in as classical gravitational waves, they will be observable inside the horizon as the universe expands as B-polarized radiation emitted from the CMB. 

\section{Conclusions}

From the non-minimal coupling of the graviton to tidal gravitational forces, QG vacuum polarization effects induce a dispersive medium with a refractive index for the propagation of gravitons. The phenomenon of gravitational birefringence occurs and in a cosmological background gravitons acquire a {\it superluminal propagation.}  The Strong Equivalence Principle is violated and this concides with the spontaneous violation of Lorentz invariance. The superluminal propagation of photons and gravitons occurs for a very short time period after the big bang. For photons it occurs at a time   $t\gtrsim 10^{-23}\,{\rm sec}$ after the big bang at an energy $\,\sim 10^5$ TeV, while for gravitons it occurs near the Planck energy at the time $t_{PL}\sim 10^{-43}\,{\rm sec}$. The calculation of QG vacuum polarization should give comparable effects to the QED vacuum polarization, inducing the refractive medium near the Planck energy $\,\sim 10^{19}$ GeV, where the gravitational fine structure constant $\alpha_g\sim {\cal O}(1)$. 

The stretching of the sub-horizon gravitational wave lengths occurs because $\lambda_g=c_g/\nu_g$ when $c_g\gg c_{g0}=c_0$.  Then, $\lambda_{gf}={\cal Q}\lambda_{g0}$ where $\lambda_{gf}$ and $\lambda_{g0}$ denote the final and initial gravitational wavelengths and ${\cal Q}\geq 10^{30}$. The wavelengths of tiny quantum perturbative density fluctuations $\delta\phi$ and primordial gravitational waves are stretched to super-horizon scale. The classical, scale invariant gravitational waves are frozen in and can be observed by B-polarization radiation emitted from the CMB. This allows for a scenario that can be compared with the stretching of tiny quantum fluctuations $\delta\phi$ and primordial gravitational waves in inflation models for which $\lambda_{gf}\propto a(t)\propto\exp(Ht)$. 

The superluminal speeds of photons and gravitons induced by quantum vacuum polarization of curved spacetime, removes the {\it ad hoc} assumption that superluminal speeds of light and gravitational waves occur in the early universe in VSL cosmology. If the B-polarization experiments do detect primordial gravitational waves this opens a new important window into the beginnings of the universe and it can reveal a possible experimental verification of quantum gravity. The superluminal propagation of gravitons and the QG vacuum polarization induced dispersive medium predicted to occur in the early universe can provide an important signal for quantum gravity.

\section*{Acknowledgments}

I thank Viktor Toth and Martin Green for helpful discussions. This research was generously supported by the John Templeton Foundation. Research at the Perimeter Institute for Theoretical Physics is supported by the Government of Canada through industry Canada and by the Province of Ontario through the Ministry of Research and Innovation (MRI).


\begin{thebibliography}{10}

\bibitem{Moffat} J.~W.~Moffat, Int. J. Mod. Phys., {\bf D2}, 351 (1993), arXiv:9211020 [gr-qc].

\bibitem{Moffat1} J. W. Moffat,  Found. Phys., {\bf 23}, 411 (1993), arXiv:9209001 [gr-qc].

\bibitem{Moffat2} J. W. Moffat, arXiv:1404.5567 [astro-ph].

\bibitem{PlanckMission1} P.~Ade et al., Planck Collaboration, arXiv:1303.5076 [astro-ph].

\bibitem{PlanckMission2} P. ~Ade et al., Planck Collaboration, arXiv:1303.5082 [astro-ph].

\bibitem{PlanckMission3} P. ~Ade et al., Planck Collaboration, arXiv:1303.5084 [astro-ph].

\bibitem{BICEP2} BICEP2 Collaboration, arXiv:1403.3985 [astro-ph].

\bibitem{Drummond} I. T. Drummond and S. Hathrell, Phys. Rev., {\bf D22}, 343 (1980).

\bibitem{Shore1} G. M. Shore, Nucl. Phys., {\bf B460}, 379 (1996), arXiv:9504041 [gr-qc].

\bibitem{Shore2} T. J. Hollowood and G. M. Shore, Int. J. Mod. Phys., {\bf D21}, 1241003 (2012), arXiv:1205.3291 [hep-th]. 
This paper contains a list of earlier published papers by the authors.

\bibitem{Weinberg} S. Weinberg, {\it Gravitation and Cosmology: Principles and Applications of the General Theory of Relativity}, John Wiley and Sons, Inc, New York, 1972.


\end{thebibliography}
\end{document}